\documentclass[10pt,aps,twocolumn,amsmath,amssymb,floatfix]{revtex4-1}
\usepackage[english]{babel}
\usepackage{graphicx}
\usepackage{amsmath}
\usepackage{dcolumn}
\usepackage{times}
\usepackage{psfrag}
\usepackage{float}
\usepackage{color}

\renewcommand{\AA}{\text{\r{A}}}

\begin{document}

\title
{Phonon-induced electronic relaxation in a strongly correlated system:  the Sn/Si(111) $(\sqrt 3 \times \sqrt 3)$ adlayer revisited}

\author
{Maedeh Zahedifar and Peter Kratzer}

\affiliation{
Fakult\"at f\"ur Physik, Universit\"at Duisburg-Essen, Campus Duisburg,
Lotharstr.~1, 47057 Duisburg, Germany
}


\begin{abstract}
The ordered adsorbate layer Sn/Si(111) $(\sqrt 3 \times \sqrt 3)$ with coverage of one third of a monolayer is considered as a realization of strong electronic correlation in surface physics. 
Our theoretical analysis shows that electron-hole pair excitations in this system can be long-lived, up to several hundred nanoseconds, since the decay into surface phonons is found to be a highly non-linear process. 
We combine first-principles calculations with help of a hybrid functional (HSE06) with modeling by a Mott-Hubbard Hamiltonian coupled to phononic degrees of freedom. The calculations show  that the  Sn/Si(111) $(\sqrt 3 \times \sqrt 3)$ surface is insulating and the two Sn-derived bands inside the substrate band gap can be described as the lower and upper Hubbard band in a  Mott-Hubbard model with $U=0.75$~eV. 
Furthermore, phonon spectra are calculated with particular emphasis on the Sn-related surface phonon modes. 
The calculations demonstrate that the adequate treatment of electronic correlations leads to a stiffening of the wagging mode of neighboring Sn atoms; thus, we predict that the onset of electronic correlations at low temperature should be observable in the phonon spectrum, too. 
The deformation potential for electron-phonon coupling is calculated for selected vibrational modes and the decay rate of an electron-hole excitation into multiple phonons is estimated, substantiating the very long lifetime of these excitations.
\end{abstract}

\maketitle

\section{Introduction}

Quantum systems defined on a lattice have received attention because they allow one to investigate quantum correlations in a well-controlled way and hence are of recent interest in the context of Quantum Simulators \cite{Cirac:2012kx,Bernien:2017uq,Zhang:2017fk}. 
In condensed matter, ordered  layers of atoms, several lattice constants apart, on an insulating surface come close to realizing a lattice system with well-defined electronic correlation. 
For instance, atoms of group-IV form ordered structures on the Si(111) surface at sub-monolayer coverage. The electronic bands inside the principal band gap of silicon formed by dangling bond orbitals of the adatoms remain  narrow if  neighboring atoms in the structure are placed several Angstroms apart. 
This narrowness of the band is a prerequisite for electronic correlation effects to become noticeable.  
From electron counting arguments, one could expect a surface band on Si(111) to be half-filled in case of group-IV adatoms (Si, Ge or Sn). 
There are two complimentary (as we will discuss below) ways of stabilizing the system that both result in a gapped electronic structure: the adatoms may undergo a Jahn-Teller-like distortion leading to a superstructure with inequivalent dangling bond orbitals, or Mott-Hubbard physics may open a gap between the filled and the unoccupied part of the electronic structure. 
Recently, not only the ground state of strongly correlated systems, but also their dynamics on ultra-short time scales has come into focus of experimental investigations \cite{Hansmann16,Ligges18}. Therefore, understanding the elementary excitations and their interplay, which determines the lifetime of excited states, has become an issue of current interest. 
The present work is intended as a first step towards studying dynamical phenomena in surface system that display strong electronic correlations.

For the system Sn/Si(111) $(\sqrt 3 \times \sqrt 3)$, the prevailing experimental and theoretical view \cite{Modesti07,Profeta07} asserts that Mott-Hubbard physics 
governs the electronic structure over the whole temperature range down to liquid helium temperatures, 
and is responsible for the formation of a gap in the electronic spectrum \cite{Hansmann13,Li13,Odobescu17}. 
This is in contrast to 
the more complex physics in 
the related system Sn/Ge(111) $(\sqrt 3 \times \sqrt 3)$ where both experiments and first-principles calculations point to a Jahn-Teller-like mechanism that stabilizes the system in a $(3 \times 3)$ superstructure with Sn adatoms at different adsorption heights 
in an intermediate temperature range between 30~K and $\sim200$~K \cite{Carpinelli:1997,Cortes:2006}.  However, more recent experimental studies found a reconstruction of Sn/Ge(111) again with the smaller $(\sqrt 3 \times \sqrt 3)$ unit cell at temperatures below 30~K that undergoes a transition to an  insulating phase upon further cooling \cite{Cortes:2013}. 
While the Jahn-Teller instability makes itself noticeable in the softening of the corresponding phonon mode showing up as height fluctuations of the surface Sn atoms \cite{Avila:1999}, the effect of strong electronic correlations on the phonon spectrum has received less attention, 
with the exception of a very recent Raman spectroscopy study \cite{Halbig:2019}. 
We will show that electronic correlations in Sn/Si(111) $(\sqrt 3 \times \sqrt 3)$ favor a planar geometry and lead to a {\em stiffening} of the associated phonon mode. 

We briefly summarize the present understanding of the physics at the Sn/Si(111) surface that has been gained from experimental studies: 
Experiments using scanning tunneling spectroscopy (STS) indicate that the  Sn/Si(111) $(\sqrt 3 \times \sqrt 3)$ system undergoes a metal-to-insulator transition below 30~K. Below this temperature, the pseudo-gap near the Fermi energy develops into a sharp gap \cite{Modesti07} whose width is reported to be 35~meV according to recent measurements 
employing scanning tunneling spectroscopy (STS)~\cite{Odobescu17}. 
Below and above the Fermi energy $E_F$, experiments using angle-resolved photoemission (ARPES) \cite{Modesti07} or inverse photoemission (KIRPES)~\cite{Charrier01} show pronounced quasiparticle peaks. These peaks have an estimated separation on the energy scale of roughly 200 meV (see also the discussion in Ref.~\onlinecite{Li13}). 
The deviating results from STS and photoemission experiments could point to the role of inelastic channels in the STS; e.g. low-energy spin modes could assist the tunneling between the surface and the probing tip, thereby masking the static band gap. Indeed, the ground state of the Sn/Si(111) surface is likely to display some magnetic ordering that can give rise to such magnetic excitations. For instance, many-body calculations suggested a non-collinear $120^{\circ}$ ordering \cite{Schuwalow10,Badrtdinov:2016} of spins at the Sn atoms.
According to  recent studies, the most likely candidate for the spin ordering on the  Sn/Si(111) $(\sqrt 3 \times \sqrt 3)$ surface is a row-wise antiferromagnetic structure \cite{Li13,Lee14} in a  $(2\sqrt 3 \times \sqrt 3)$ supercell. 
This is supported by recent experimental studies using spin-selective photoemission \cite{Jaeger:2018}, and forms the basis for the present study.

\begin{figure}[tbh]
\begin{center}
\includegraphics[width=0.45\textwidth]{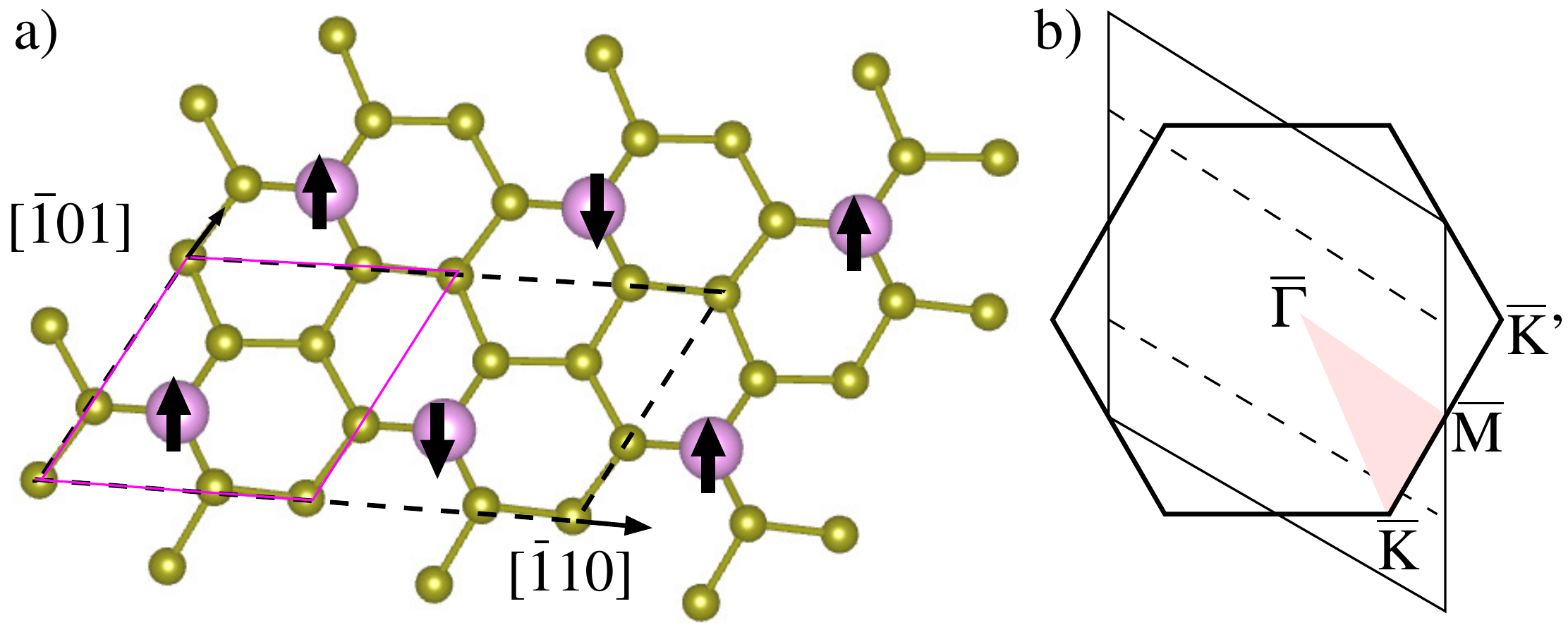} 
\caption{
a) Top view of Sn/Si(111). Si atoms are shown in yellow, Sn atoms as  pink balls. The $\sqrt{3}\times \sqrt{3}$ and $2\sqrt{3}\times \sqrt{3}$ unit cells for antiparallel spin arrangement are indicated by the full  pink and the dashed black polygon.  
b) Brillouin zone of $\sqrt{3}\times \sqrt{3}$ (full lines, rhombic shape or hexagon) and $2\sqrt{3}\times \sqrt{3}$ (dashed lines). 
In reciprocal lattice coordinates, we have $\bar{\rm M}=(0,1/2), \bar{\rm K}=(-1/3,1/3),\bar{\rm K'}=(1/3,2/3)$. 
For band plotting later on, the k-point path marked by the filled triangle $\bar \Gamma - \bar {\rm K} - \bar {\rm M} - \bar \Gamma$ is used.}
\label{fig:geom}
\end{center}
\end{figure}

In this work, 
our main interest is in the lifetime of electronically excited states in Mott-Hubbard systems, exemplified by the Sn/Si(111) surface. In two-photon pump-probe spectroscopy, it should be possible to observe the lifetime of electrons in the upper Hubbard band by a time-delayed probe pulse after creating an excited state of the electronic system by a suitable pump pulse, possibly after allowing for a very short electronic relaxation that populates the upper Hubbard band. 
Since the Mott-Hubbard gap, at least according to  the ARPES \cite{Modesti07} and KIRPES~\cite{Charrier01} experiments, is supposed to be quite large, the excitation can only relax by coupling to a high number of phonons.  
To estimate the relaxation rate, calculations of the phonon spectrum and of electron-phonon coupling constants are required. In addition, we search for fingerprints of the electronically correlated ground state in the phonon spectra. 
We perform DFT calculations using the semi-local PBE functional to obtain the surface phonon spectrum. These are complemented by hybrid-functional calculations of the electronic band structure and the electron-phonon coupling for selected phonon modes with out-of-phase motions of the Sn atoms. 
We find that the magnetic ground state found with the hybrid functional calculations is accompanied by a stiffening (compared to semi-local DFT) of the phonon mode corresponding to the 'wagging' motion of the Sn atoms. Thus, experimental probes sensitive to surface vibrations in the low-frequency range below two THz, e.g. Raman spectroscopy or helium atom scattering, could provide useful hints about the nature of the electronic ground state of this system. Extending over previous theoretical work based on band structures obtained with (semi-)local density functionals, 
we show that the data extracted from the DFT calculations can be used as input to a model Hamiltonian that allows us to estimate the lifetime of excitons due to phonon emission in a non-linear process. While the electron-phonon coupling is found to be sizable, the decay of the exciton is exponentially suppressed due to the very large number of phonons involved in this process.  
Therefore, we suggest two-photon pump-probe experiments to be carried out for Sn/Si(111). Since the excited state is predicted to be sufficiently long-lived, such an experiment could provide information not only about the valence bands, but also about the dispersion of the unoccupied bands and their relative position with respect to the valence band maximum. Moreover, detecting the electrons both from single-photon as well as two-photon photoemission from the same sample could resolve ambiguities about the size of the band gap in this much-studied system, and provide useful data to be compared with many-body calculations~\cite{Schuwalow10,Li13,Hansmann16} of the band structure.

\section{Methods}
A detailed understanding of the electronic structure of correlated systems requires special methods, for instance the Dynamical Cluster Approximation   applied in Ref.~\onlinecite{Li13}. 
However, these methods, as well as the calculation of phonon spectra, rely on input from density functional theory (DFT). If applied with care, DFT calculations can already reveal salient features and allow us to address the various energy scales in the problem. 
While local or semi-local density functionals find the  Sn/Si(111) $(\sqrt 3 \times \sqrt 3)$ to be metallic, the HSE hybrid functional \cite{Heyd2003,Heyd2006} correctly reproduces the insulating ground state\cite{Lee14}. Technically speaking, electronic correlations are treated in the HSE functional as part of the (screened) electronic exchange. 
Therefore, despite their failure to yield the proper electronic excitation spectrum,  HSE calculations can still be used to estimate the basic parameters (on-site Coulomb repulsion, exchange-coupling and electron-phonon coupling parameters) that are required as input to model Hamiltonians (cf. Ref.~\onlinecite{Lenarcic2015}) employed to describe the electronic relaxation. 
However, in our understanding, the opening of a gap in the HSE calculation should not be taken as evidence of the system being a Slater insulator, as claimed in previous work \cite{Lee14}, due to the ambiguity in the assignment of many-particle interactions to the exchange or correlation part of the potential in a hybrid functional.

In our numerical modeling, the Sn/Si(111) surface with a coverage of 1/3 is described by a repeated slab geometry.  
The supercell was built with 10 $\rm{\AA}$ of vacuum in between the slabs. 
The Si lattice constant $a_{0}$ = 5.445 (5.468) $\rm{\AA}$ for the HSE (PBE) calculation was used. 
It is important for the phonon calculation performed later on that the consistent lattice constant for each functional is employed to avoid any stress in the crystallographic ground state. For bulk silicon, a consistent treatment using the HSE functional has been shown to give excellent results for the phonon frequencies \cite{HuHa09}. To assess the role of magnetic ordering, collinear spin-polarized calculations were carried out in addition to the standard  non-spinpolarized calculation.  A doubled unit cell  with lateral $2\sqrt{3}\times \sqrt{3}$ periodicity was used to allow for antiparallel alignment of the magnetic moments at the two Sn atoms. The top view of the slab with both  $\sqrt{3}\times \sqrt{3}$ and $2\sqrt{3}\times \sqrt{3}$ unit cells is displayed in Fig.~\ref{fig:geom}.  
In order to sample the Brillouin zone of the $\sqrt{3}\times \sqrt{3}$ and $2\sqrt{3}\times \sqrt{3}$ unit cells, we used a 8$\times$8$\times$1 and 4$\times$8$\times$1 Monkhorst-Pack \cite{MoPa76} k-point mesh, respectively. 
However, all band structures shown later on are still plotted along the conventional k-point path $\bar \Gamma - \bar {\rm K} - \bar {\rm M} - \bar \Gamma$ of the $\sqrt{3}\times \sqrt{3}$ Brillouin zone. The Si(111) slab consisted of five bilayers of Si, and
the Si dangling bonds on the rear side of the slab were saturated by hydrogen atoms while one Sn atom in $\sqrt{3}\times \sqrt{3}$ (two tin atoms in $2\sqrt{3}\times \sqrt{3}$) were placed on the front surface in the $T_4$ site. The position of Sn atoms, as well as of the Si atoms in the first eight layers, were relaxed, while the two layers of Si atoms at the rear side are held fixed together with the hydrogen atoms. The residual force components were less than 0.01 eV/$\rm{\AA}$. 
Some calculations with similar setting were performed for Sn atoms on a Ge(111) slab for comparison.

Calculation of phonons and electron-phonon coupling of this system are interesting for us, so the {\sc Phonopy}~\cite{phonopy} package in conjunction with the {\sc VASP}~\cite{VASP1,VASP2} code was used for calculating the phonon band dispersion.
Finite atomic displacements were used together with a large supercell obtained by doubling the size of the initial supercell in both lateral directions. 
Atomic forces were calculated for 48 displaced configurations in case of the $\sqrt{3}\times \sqrt{3}$ structure.

For the electronic band structure, density functional theory calculations were performed by using the {\sc FHI-AIMS} \cite{FHIaims,FHIaimsHSE} code, which is an accurate all-electron full-potential electronic structure package based on numeric atom-centered orbitals, with so-called 'tight' computational settings. 
In this part of the work, the screened hybrid functional HSE06 with the mixing factor $\alpha = 0.25$ and screening parameter $\omega = 0.11$ bohr$^{-1}$ was used for the exchange-correlation energy. 
A slab geometry with the HSE lattice constant of  bulk Si was constructed, and the atomic positions in the electronic ground state were relaxed. 

\begin{figure}[tbh]
\begin{center}
\includegraphics[width=0.45\textwidth]{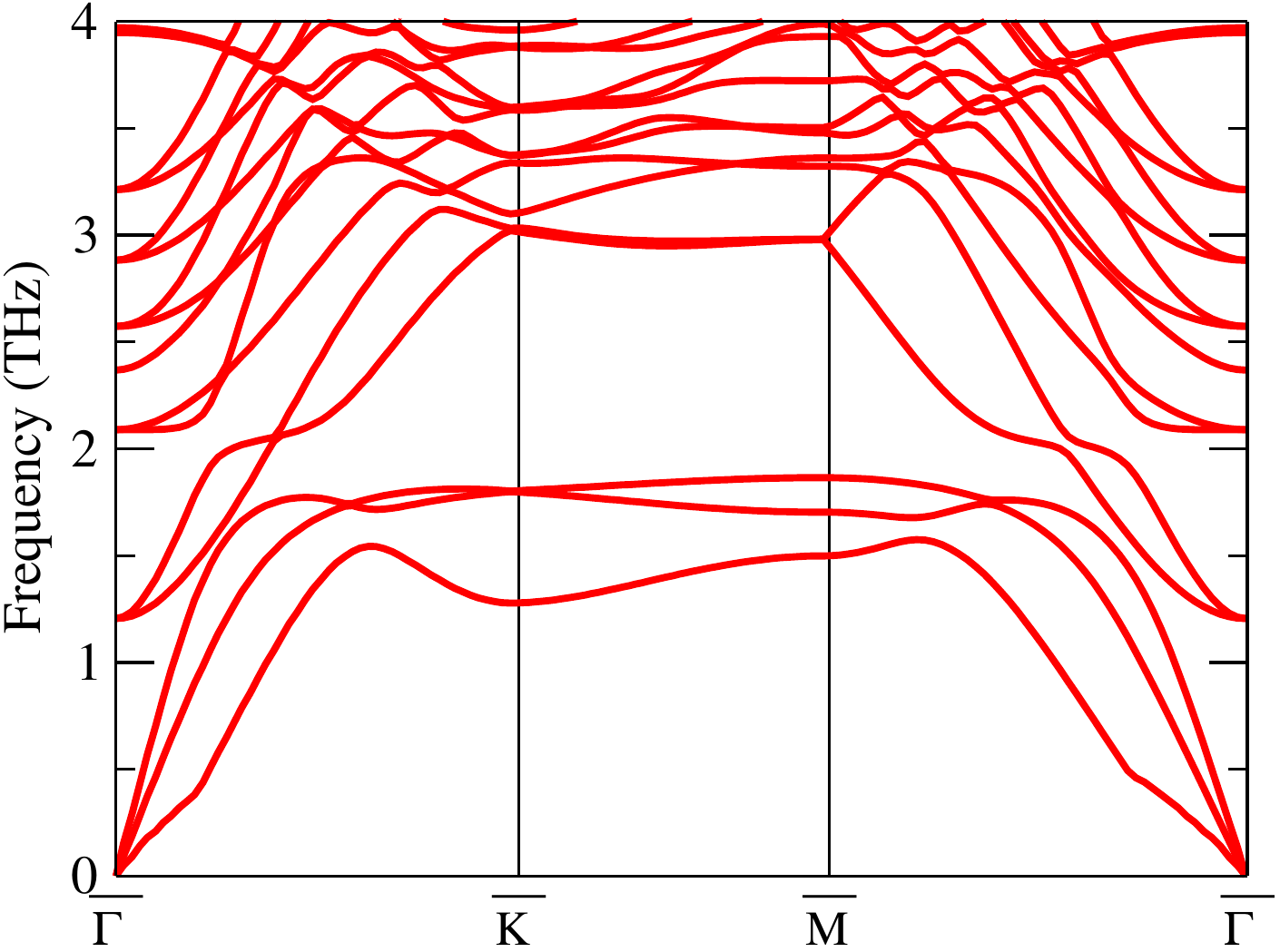} 
\caption{Phonon dispersion for $\sqrt{3}\times\sqrt{3}$ Sn/Si(111) obtained with the PBE functional in a non-spinpolarized calculation.}
\label{fig:phon}
\end{center}
\end{figure}

\section{Results}

\subsection{Phonon and electron band structure from GGA-PBE calculations}
We start by  confirming previous calculations that had demonstrated that a non-spinpolarized PBE calculation of the Sn/Si(111)$(\sqrt 3 \times \sqrt 3)$ leads to a metallic surface band (See Supplemental Information). 
Despite the lack of proper electronic correlations in this approximative method, it is instructive to calculate the phonon spectrum to get a first idea of the surface vibrational properties.     
Phonon spectra $\omega_i(q)$ obtained with the PBE functional in the non-magnetic (NM) ground state with the finite-displacement method implemented in {\sc Phonopy} are presented in Fig.~\ref{fig:phon}.
The phonon modes related to Sn appear in two groups: 
There is one mode in which both the Sn atom and the top-most layer of Si atoms move in phase ($\omega(q=0) = 4.00\,$THz). 
Lower-lying modes involve out-of-phase motions of the Sn atom and the neighboring Si atoms. These comprise motions of the Sn atom within the surface plane, leading to two bands with small dispersion ($\omega = 1.75\,$THz at $\bar {\rm K}$ in the NM calculation), and a more dispersive band corresponding to a vertical motion of the Sn atom relative to the Si atoms ($\omega = 1.30\,$THz at $\bar {\rm K}$ in the NM calculation). The latter mode shows a slight softening at the edge of the Brillouin zone at the point $\bar {\rm K}$. 
Here, Sn atoms in neighboring unit cells are moving with opposite phase. These Sn atoms are connected by a Si--Si bond lying in the surface plane. The whole group of atoms, Sn--Si--Sn--Sn, performs a 'wagging' motion.  
In agreement with previous studies~\cite{Flores00}, we interpret the mode softening as a first indication of a possible instability of the (metallic) PBE band structure with respect to Jahn-Teller distortions. 
However, this finding is {\em not} confirmed by our HSE hybrid functional calculations (see below). 

\begin{figure}[tbh]
\begin{center}
\includegraphics[width=0.45\textwidth]{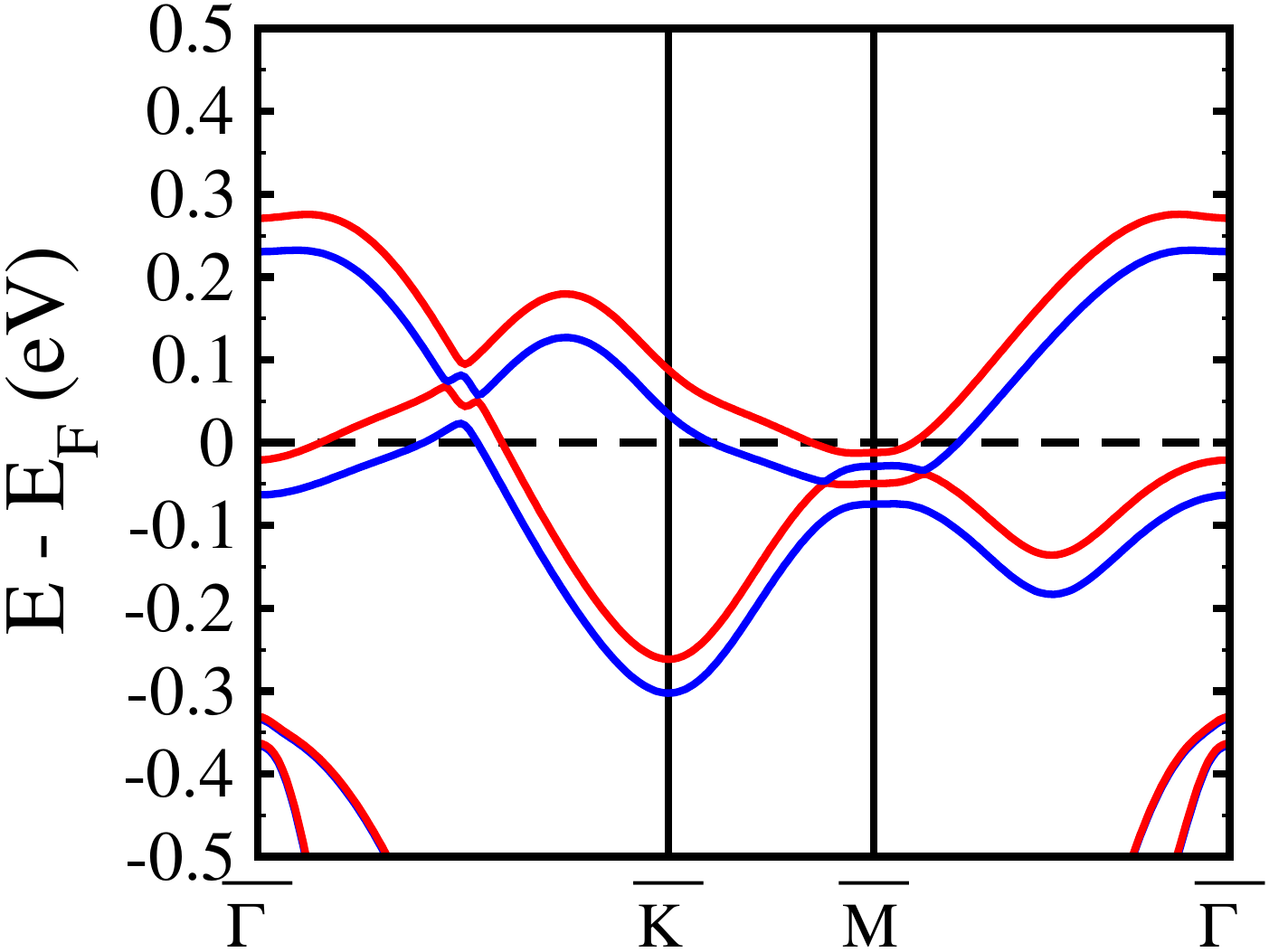} 
\caption{The electronic band structure for AFM spin ordering obtained with the PBE functional including spin-orbit coupling. 
Calculations are performed in a  $(2 \sqrt 3 \times \sqrt 3)$ unit cell with  two Sn atoms.  The red 
(mostly spin-up) and blue 
(mostly spin-down) bands are related to different superpositions of spin wave functions.}
\label{fig:elBandStructPBE}
\end{center}
\end{figure}

In order to investigate to which extent the AFM spin ordering can already account for the correlated electronic features, we performed PBE calculations with the $(2 \sqrt 3 \times \sqrt 3)$ unit cell where the magnetic moments of the Sn atoms were frozen in the AFM state. 
In these calculations, the energy of the AFM configurations was found to be lower than the energy of the  non-spinpolarized calculation by 6~meV per Sn atom on Si(111). In the electronic band structure, the occupied and the unoccupied surface states still overlap in energy, yielding a metallic surface, as seen in Fig.~\ref{fig:elBandStructPBE}. However, the band width of 275~meV of the unoccupied band along $\bar \Gamma \bar {\rm M}$ is slightly reduced relative to the non-spinpolarized PBE calculation (See Supplemental Information).

\begin{figure}[ptbh]
\begin{center}
\includegraphics[width=0.50\textwidth]{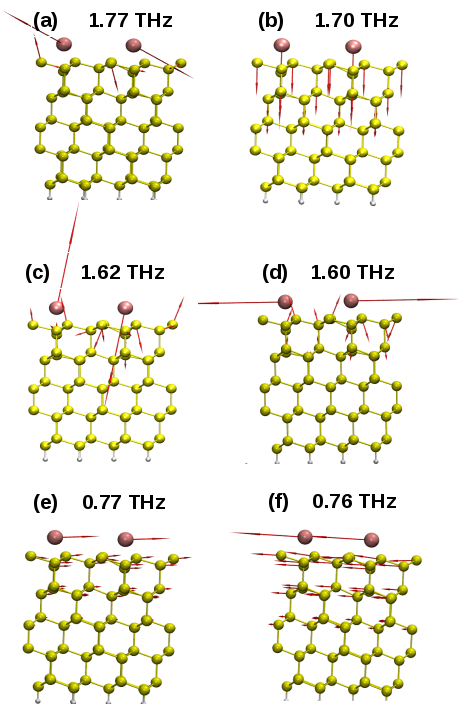} 
\caption{ Side view of Sn (pink spheres) on Si(111)
(yellow spheres). Atomic displacements for selected phonon modes with $q=0$ of
Sn/Si(111) $(2\sqrt{3}\times \sqrt{3})$ are indicated by the arrows. Calculation are performed with PBE for the AFM state. The  frequencies (energies) of the modes shown in  (a), (b), (c), (d), (e) and (f) at $q=0$ are 1.77~THz (7.35~meV), 1.70~THz (7.04~meV), 1.62~THz (6.71~meV), 1.60~THz (6.65~meV), 0.77~THz and 0.76~THz (both 3.16~meV), respectively. (a) and (c) are the 'torsion' and the 'wagging' mode, respectively.}
\label{fig:eig}
\end{center}
\end{figure}

The calculations of phonon frequencies with the PBE functional in the AFM ground state were repeated 
in the $(2 \sqrt 3 \times \sqrt 3)$ unit cell.  
Due to the smaller Brillouin zone compared to the $(\sqrt 3 \times \sqrt 3)$ unit cell, backfolding of the phonon modes onto the $\bar \Gamma$ point is presented 
as discussed in the Supplemental Material. 
Since phonon band structure calculations with a very large number of atoms in the unit cell are susceptible to small errors in the forces on the atoms, we take less cumbersome and more accurate calculations carried out separately for $q =0$ as the basis of the following discussion. 
The phonon frequencies and corresponding eigenmodes are displayed in Fig. \ref{fig:eig}. 
Interestingly, the 'wagging' mode of the Sn atoms has now moved to higher frequency $\omega(q=0) =1.62\,$THz (Fig. \ref{fig:eig}(c)). 
For comparison with the NM calculation, taking into account backfolding due to the different Brillouin zone, one should use the frequency of the corresponding mode at the $\bar {\rm K}$ point, $\omega(q=0) =1.30\,$THz.
Thus, taking into account the 
AFM magnetic order at the Sn atoms results in a stiffening of this mode. 
Freezing the spin density at the Sn atoms prevents the Jahn-Teller effect previously predicted for the metallic, non-magnetic surface. 
In addition, there is a slightly higher mode at $\omega(q=0) =1.77\,$THz involving opposite motion of the Sn atoms (Fig. \ref{fig:eig}(a)). 
By analyzing the mode eigenvector, it can be characterized by a torsion of the Sn--Si--Si-Sn groups mainly within the plane of the surface.
Again, in the NM calculation this should be compared to the mode at the $\bar {\rm K}$ point, $\omega(q=0) =1.75\,$THz. Thus, 
its frequency is almost unchanged; i.e. it is insensitive to the magnetic correlations at the Sn atoms.

As a side remark, it is interesting to compare these findings to the situation on the Sn/Ge(111) surface. Our PBE calculations confirm that the potential energy surface has a very flat minimum with respect to alternating upward and downward displacements of the Sn atoms, as found in previous publications \cite{Avila:1999,Flores00}. This allows for height fluctuations in the intermediate temperature range, consistent with the interpretation of the experimental findings. 
In addition, our non-spinpolarized PBE calculations show a mode characterized by lateral motion of the Sn atom at 1.67~THz, in agreement with Ref.~\onlinecite{Halbig:2019}. 
Remarkably, we find both the FM and the AFM states in a spin-polarized PBE calculation with a $(2 \sqrt 3 \times \sqrt 3)$ unit cell to be energetically higher than the NM state, by 10 and 11 meV per Sn atom, respectively. 
Because of the instability of the magnetic states, we did not further investigate the role of magnetic ordering on the phonon spectrum in this system.

\subsection{Hybrid functional calculations of electronic band structure and electron-phonon coupling}

For the calculations with the hybrid functional HSE, a larger $(2 \sqrt 3 \times \sqrt 3)$ unit cell with two Sn atoms was used. The ground state is found to be a spin-polarized state in which the magnetic moments at the two Sn atoms are pointing in opposite directions, i.e. they show row-wise antiferromagnetic (AFM) ordering. 

\begin{figure}[tbh]
\begin{center}
\includegraphics[width=0.45\textwidth]{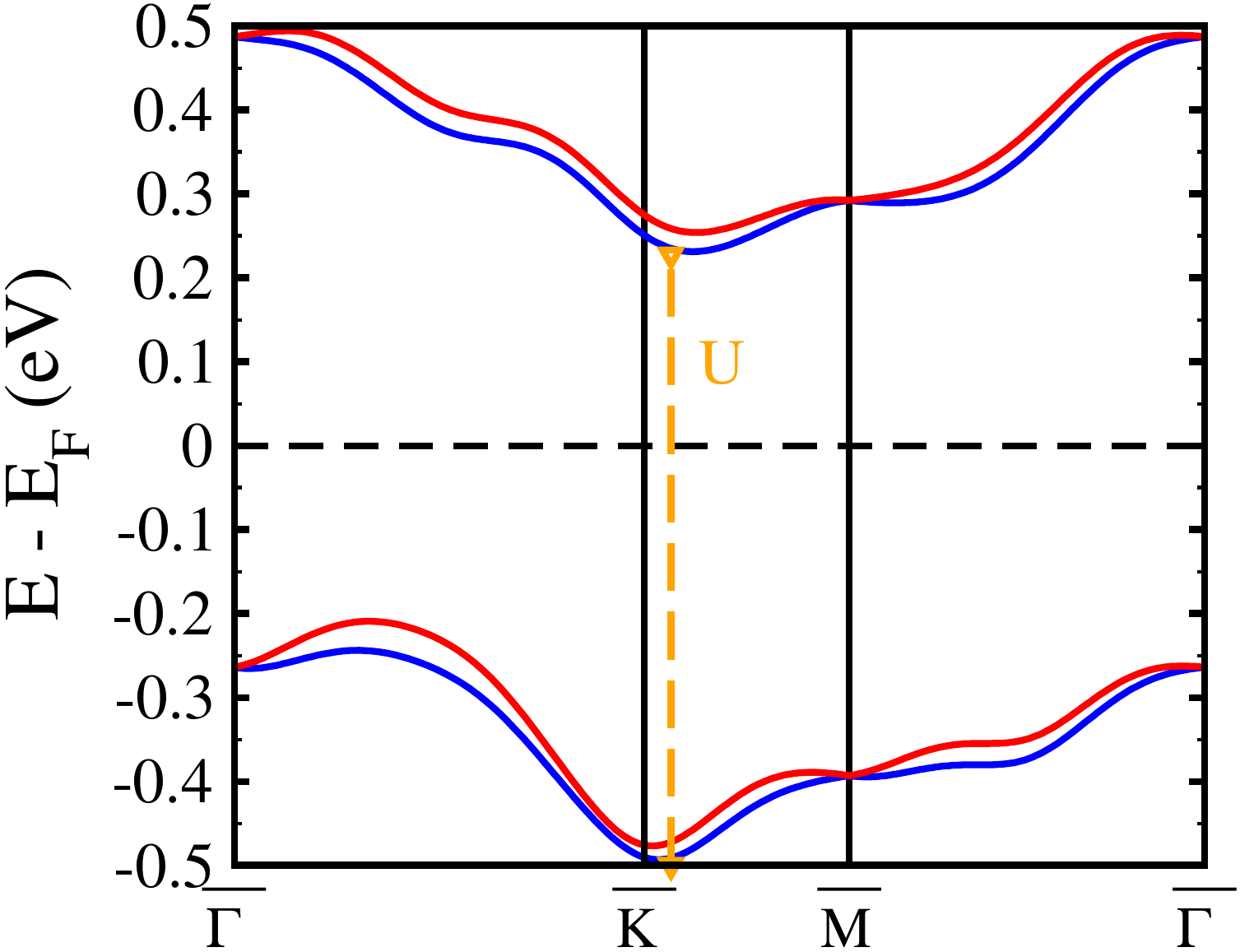}
\caption{The electronic band structure for AFM spin ordering obtained with the HSE06 functional including spin-orbit coupling. 
Calculations are performed in a  $(2 \sqrt 3 \times \sqrt 3)$ unit cell with  two Sn atoms. The red and blue bands are related to (mostly) spin-up and spin-down wave functions.}
\label{fig:elBandStructHSE}
\end{center}
\end{figure}

The electronic band structure resulting from the HSE calculation is shown in Fig. \ref{fig:elBandStructHSE}. 
It displays an indirect band gap of 540~meV including spin-orbit coupling (SOC) and 430~meV if SOC is neglected. 
This value is in line with similar theoretical work~\cite{Lee14} reporting a gap of 328~meV, but is much larger than the experimental value of 35~meV obtained from STS \cite{Modesti07}. Instead, it is closer to the value of $\sim 200$~meV obtained from ARPES~\cite{Li13}.
The width of the occupied (unoccupied) band along $\bar \Gamma \bar {\rm M}$ is $w=140$~meV ($w=200$~meV); i.e. smaller than in the PBE ground state. 
The ferromagnetic solution is $E_{\rm FM} -E_{\rm AFM}  = 25$~meV per Sn atom higher than the AFM ground state,  
i.e., the preference of AFM coupling is much enhanced compared to the calculations with the PBE functional.
Calculations including spin-orbit coupling show a splitting of the occupied surface band being largest at the $\bar {\rm K}$ point ($E_{\rm SOC} =22$~meV). 
A comparable value ($E_{\rm SOC} =36$~meV) has been reported~\cite{Badrtdinov:2016} from PBE calculations using a FM ground state.

\begin{figure}[tbh]
\begin{center}
\includegraphics[width=0.40\textwidth]{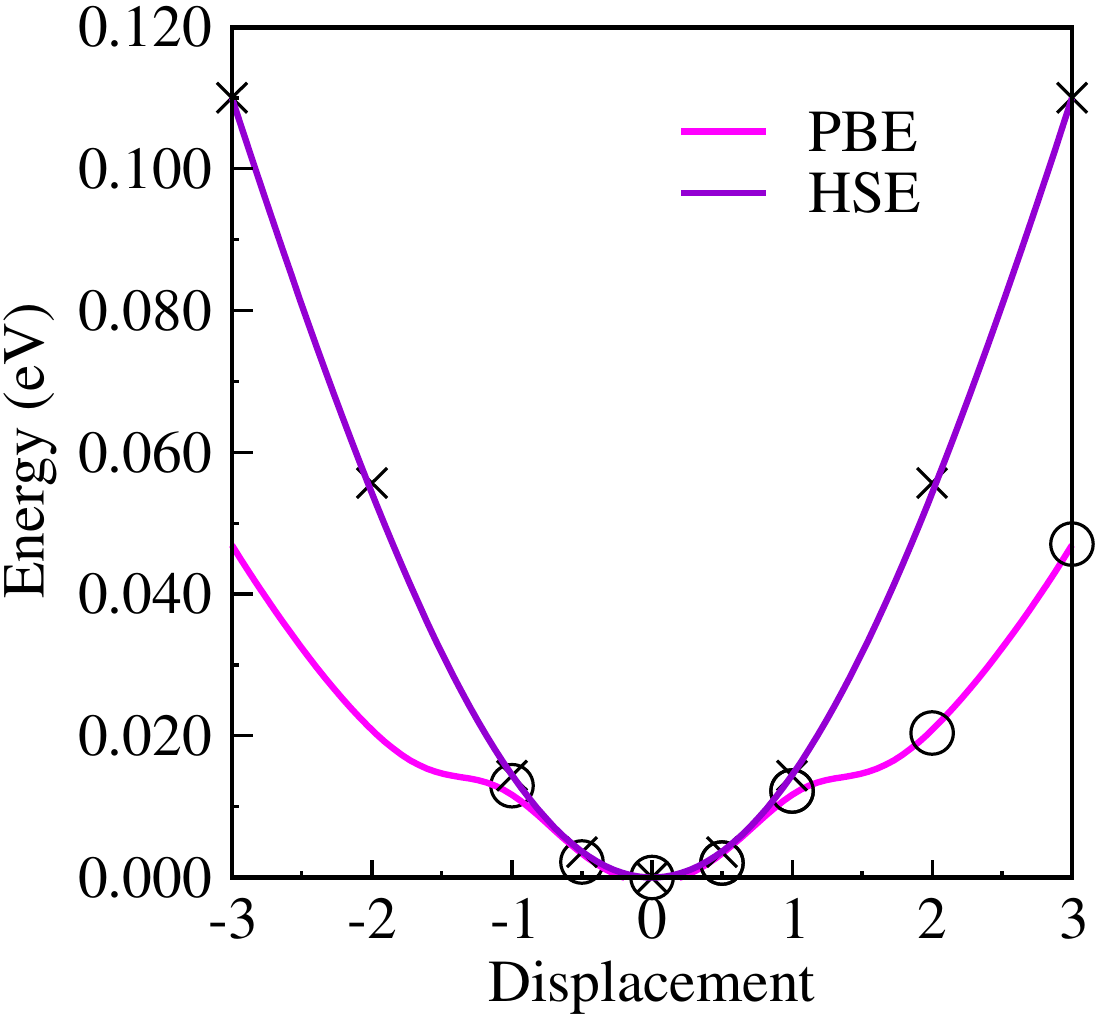} 
\caption{Energy as function of displacement along the wagging mode at $q=0$ for (2$\sqrt{3}\times\sqrt{3})$ with the AFM ordering of the electron spins taken into account. The displacement is given in units of the oscillator length. }
\label{fig:totEnergyPhonons}
\end{center}
\end{figure}

For the 'wagging' mode of the Sn atoms, 
we performed total-energy calculations with finite displacements of both the Sn and Si atoms along the mode eigenvectors.
The results for displacements of varying magnitude are shown in Fig.~\ref{fig:totEnergyPhonons}. 
While the HSE data points fall onto a parabola, as expected for a harmonic vibration, the PBE curve shows strongly anharmonic behavior. 
The energy versus displacement curve consists of two pieces: 
For small displacements, a stiffening of the mode is observed, while the restoring forces are considerably smaller at larger displacements. 
This means that the magnetic correlations can be destroyed by large displacements in the PBE calculations, while this is not possible in the HSE calculations due to the opening of a band gap. 
Weak restoring forces after break-down of the electronic correlations cause the Sn atoms to spend relatively long time near the turning points of their oscillatory motion. 
This could explain why in photoelectron spectra \cite{Uhrberg00} of the surface recorded at a temperature of 70K two distinct peaks associated with two Sn atoms at different adsorption height have been observed.
By fitting the HSE data points, the phonon energy $\hbar\omega_{\rm HSE} = 11$~meV ($\omega_{\rm HSE} = 2.65 \, $THz) is obtained. 
A summary of the phonon frequencies obtained with different functionals of both the wagging and the torsional mode is given in Table \ref{tab:stiffening}. 
The stiffening of the 'wagging' mode observed already in the AFM spin-polarized PBE calculations is even more pronounced in the HSE calculations. 
Due to the gap in the HSE band structure, a transfer of spin density between the two Sn atoms is no longer possible. 
Thus, in the correlated electronic state, the geometrical structure in the AFM state is no longer prone to Jahn-Teller instability. 
In conclusion, stabilization by spin-ordering and displacive rearrangements of the atoms must be considered as competing stabilization mechanisms excluding each other. 
For the complementary system Sn/Ge(111), our calculated results are in line with this conclusion: Here, the spin-polarized states are found to be energetically higher than the non-spinpolarized state, which leaves only the displacive rearrangement of the atoms as stabilization mechanism. 
It is noteworthy that a recent Raman study \cite{Halbig:2019} observed peaks with $E$ symmetry at 2.65~THz for Sn/Si and  1.67~THz for Sn/Ge(111), in addition to some more noisy structures at lower frequencies. Our computed results are in agreement with both experimental values.  Moreover, 
it was reported that the 2.65~THz peak on Sn/Si(111) showed a slight stiffening when cooling from room temperature to 40~K, while such a temperature dependence was absent for Sn/Ge(111). Thus, these experiments may already reach the sensitivity to detect fingerprints of magnetic ordering in the vibrational spectra. 

\begin{figure}[ptbh]
\begin{center}
\includegraphics[width=0.4\textwidth]{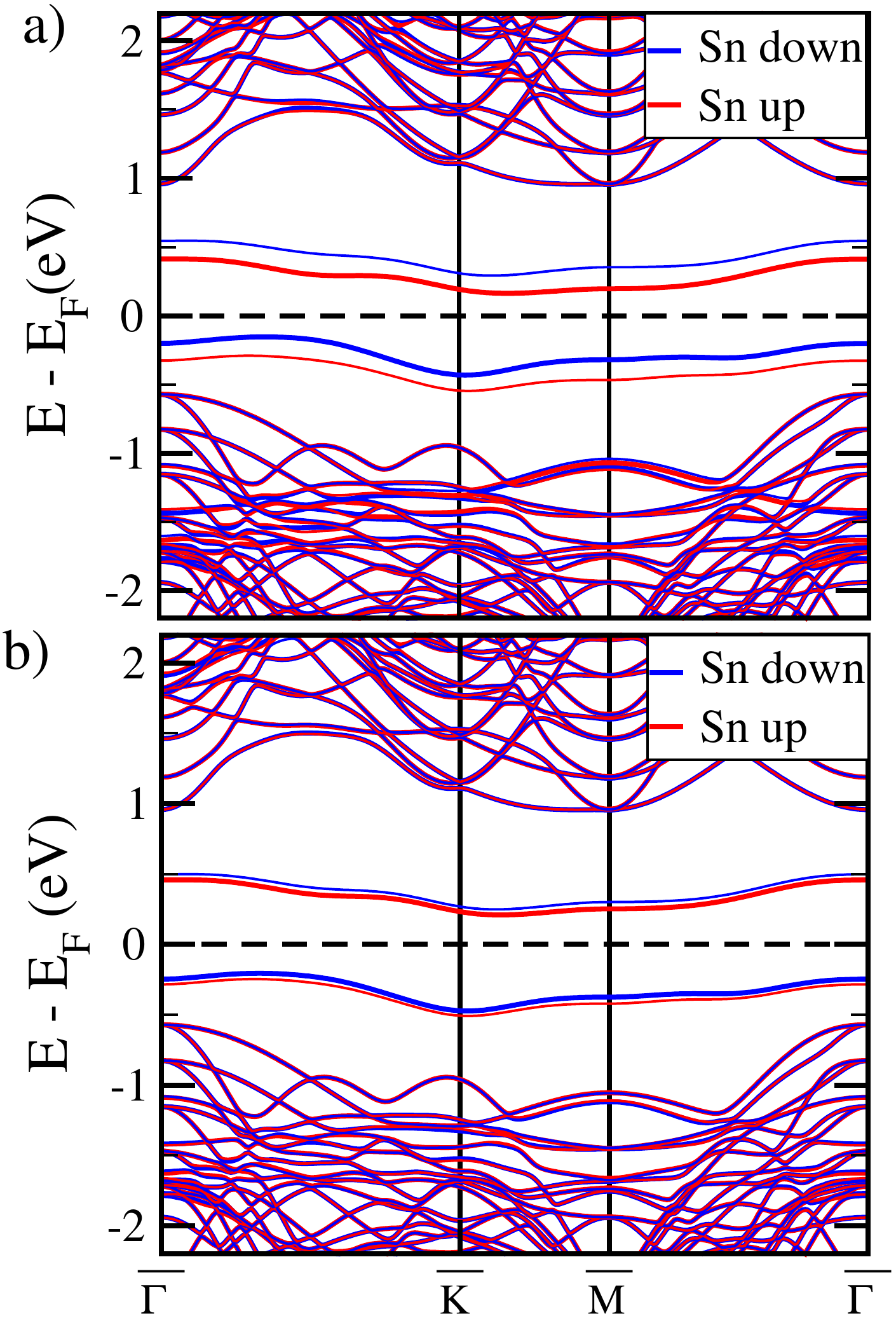}
\caption{Electronic band structure of Sn/Si(111) ($2\sqrt{3}\times\sqrt{3}$) calculated with the HSE functional for displaced structures. The displacements for a) the wagging mode and b) the torsional mode are taken from Fig.~\ref{fig:eig}. By comparing to the band positions to the undistorted structure, Fig.~\ref{fig:elBandStructHSE}, the electron-phonon coupling parameter $\lambda$ can be extracted. 
\label{fig:ephSn111} }
\end{center}
\end{figure}

While the insulating ground state of the Sn/Si(111) surface as predicted by the HSE calculations does not allow for electron-phonon {\em scattering}, it is still possible to estimate the {\em coupling} between structural deformations (a 'frozen' phonon) and the electronic band energies.
While the mode associated to the Jahn-Teller instability became stiffer in the correlated system, we still expect a sizable contribution to electron-phonon coupling from this mode. Therefore, 
for the phonon modes that involve substantial displacements of the Sn atoms normal to the surface plane in opposite directions (Fig.~\ref{fig:eig} (a) and (c)), we carry out HSE calculations of the electronic structure for various displaced atomic geometries.  
The corresponding eigenvector of the dynamical matrix, scaled with the dimensionless atomic mass factor 
$\sqrt{m_j}$ 
($j$ stands for Sn or Si, $m_{\rm Sn} = 109$amu,  $m_{\rm Sn} = 28$amu), is used to generate the atomic displacements.  
Fig.~\ref{fig:ephSn111} a) and b) show the electronic band structure after the displacements have been applied. 
It is seen that each of the two surface bands splits up into two subbands. The upper (blue) subband has its wavefunction amplitude mostly at the Sn atom that is geometrically closer to the surface plane ('down' atom), while the lower (red) band  has its wavefunction amplitude mostly at the Sn atom higher above the surface ('up' atom). 
The origin of this effect can be traced back to changes in orbital hybridization caused by the displacements: 
The surface state has a substantial contribution from the Sn $p_z$ orbital pointing perpendicular to the surface plane. 
The Sn 'up' atom is in a pyramidal coordination 
bonded to Si by its $p$ orbitals;
hence the dangling orbital has an increased $s$ orbital character and moves down in energy.  
The Sn 'down' atom moves towards the surface plane and comes closer to $sp^2$ hybridization; hence the $p_z$ character of the dangling orbital becomes enhanced, and it moves upward in energy.
Displacements along the 'wagging' mode involve larger components perpendicular to the surface, and thus have a much stronger effect on the band positions, compared to the 'torsion' mode.

Using deformation potential theory~\cite{Resta}, one can determine quantitatively the 
electron-phonon coupling parameter $\lambda$ from the energetic shifts (relative to the ground state) of the bands caused by the displacements. Specifically, we calculate 
$$
\lambda = \left( \frac{\hbar}{2 m_{\rm Sn} \omega_{\rm HSE}} \right)^{1/2} \frac{\Delta \varepsilon}{\Delta x} = \frac{\Delta \varepsilon}{\sqrt 2}
$$
where $\Delta \varepsilon$ is the energy shift of the highest occupied band in the HSE calculation and $\Delta x$ is a  displacement of one oscillator length.  
Using the band energies at the the $\bar {\rm K}$ point, we obtain from Fig.~\ref{fig:ephSn111} the values $\lambda = 49$~meV and $\lambda = 13$~meV 
for the 'wagging' and the 'torsion' mode, respectively. 

\begin{table}[tbh]
\caption{Vibrational frequencies in THz of the modes with opposite displacements of the Sn atoms, obtained with the PBE or the HSE functional. 
NM and AFM stand for non-magnetic calculations and calculations with antiferromagnetic order of the magnetic moments at the Sn atoms, respectively. 
NM frequencies were taken at the $\bar {\rm K}$ point in Fig.~\ref{fig:phon}, AFM frequencies at the $\bar \Gamma$ point.
Considerable mode stiffening, in particular of the wagging mode, is observed when AFM spin correlations are taken into account.}
\begin{center}
\begin{tabular}{l | c | c | c}
               & PBE (NM) & PBE (AFM) & HSE (AFM) \\
\hline
\hline 
wagging & 1.30 & 1.62 & 2.65 \\
torsional & 1.75 & 1.77 & 2.08 \\
\end{tabular}
\end{center}
\label{tab:stiffening}
\end{table}

\subsection{Implications for the lifetime of excited states}

The electronic and phononic structure data obtained by our DFT calculation can be used to obtain a better understanding of relaxation processes in this strongly correlated system. 
Optical excitation of the surface will generate electron-hole pairs. 
Recently, the various processes contributing to the decay of such an excitation have been studied analytically, employing an effective Mott-Hubbard (MH) Hamiltonian coupled linearly to a phonon bath \cite{Lenarcic2015}. 
To make contact to this study,  it is helpful to cast our DFT results into the form of an effective MH Hamiltonian  
\begin{eqnarray}
H &=& - t \sum_{\langle i,j \rangle} (\hat{c}^{\dag}_{i\sigma} \hat{c}_{j\sigma} + \mathrm{h.c.}) + U \sum_{i} \left(\hat{n}_{i\uparrow}- {1 \over 2} \right) \left(\hat{n}_{i\downarrow}- {1 \over 2} \right) \nonumber \\
& &  + {1 \over 2} V \sum_{\langle ij \rangle \sigma \sigma'} \left(\hat{n}_{i\sigma}- {1 \over 2} \right) \left(\hat{n}_{j \sigma'}- {1 \over 2} \right) \, ,
\end{eqnarray}
where Coulomb interaction between pairs of neighboring sites, denoted by $\langle ij \rangle$, has been taken into account via an off-site interaction $V$ in addition to the usual on-site interaction $U$. 
The hopping matrix element $t$ can be estimated from the band width of the DFT calculations 
with the HSE functional,
yielding $t= w/8 = 25$~meV.
The order of magnitude agrees 
with previous estimates of the hopping matrix element~\cite{Li13} 
using the local density approximation, but the HSE calculation yields a band width smaller by a factor two.
We stress that the on-site repulsion $U$ does not allow for a simple interpretation as Coulomb repulsion between two electrons in the Sn $p_z$ orbital. Both our HSE calculations and previous work with the HSE functional \cite{Lee14} as well as with the PBE functional using Wannier functions \cite{Badrtdinov:2016} show that the electrons occupying the lower surface band are still localized, but extend over both the Sn atom and its Si neighbors. Keeping this caveat in mind, the MH Hamiltonian can be used to study the excitation spectrum.

Low-lying electronic excitations are given by AFM spin waves. 
At half filling, every site is occupied by one electron, and the simplest MH Hamiltonian with $V=0$ can be approximately transformed onto a Heisenberg model with the exchange coupling constant $J$.
The presence of spin-orbit interaction requires the generalization to an anisotropic Heisenberg model 
in which the excitation of spin waves requires a minimum energy. 
A detailled discussion of a suitable spin Hamiltonian and of its consequences for the spin excitation spectrum are found in Ref. \onlinecite{Badrtdinov:2016}. We speculate that this spin-wave gap in the spin wave spectrum may be responsible for the experimentally observed opening of a tiny gap at low temperatures in the STS experiments \cite{Modesti07,Odobescu17}, but do not attempt to estimate it quantitatively. 
 
The charged excitations are given by the addition or removal of one electron. In the hybrid functional approach, the energetic shift between conduction and valence band can be taken as a first approximation of the typical energy of such an excitation.  
This can be made clear in the following way: 
When modeling the system by a MH Hamiltonian, in the limit of 
small $V$ and large $U$, the lower and upper Hubbard band can be approximated by
$$
\varepsilon_{\pm}(k) = \frac{1}{2} \left( \varepsilon(k) \pm \sqrt{ \varepsilon^2(k) + U^2 } \right) \, .
$$
For the case $\varepsilon(k) \ll U$, the lower and upper Hubbard band are located at $\varepsilon_{\pm}  = (\varepsilon(k) \pm U)/2$, respectively. 
Thus, the energy of a charged excitation is of the order of $U$. As an estimate, we can identify $U$ with the gap between the two surface bands in the HSE calculation. Taking the direct gap at the $\bar {\rm K}$ point in the Brillouin zone in  Fig.~\ref{fig:elBandStructHSE} yields $U = 0.75$eV.
This value is in good agreement with Ref. \onlinecite{Li13} who obtained $U=0.66$ eV from a comparison between the calculated and the experimental electronic density of states. Moreover, our result is similar to the work of Hansmann {\em et al.} \cite{Hansmann13} who described the system by combining on-site and off-site interactions of $U=1.0$ eV and $V=0.5$ eV. 

If we disregard the Coulomb interaction between electron and hole, assuming that it is screened, the excitonic energy scale is the same as for charged excitations, i.e., of the order of a few tenth of eV. 
In principle, these excitations can decay by dissipating their energy to spin waves or to phonons. The relative importance of both processes depends on the spin structure of the excited many-particle states, as discussed in Ref.~\onlinecite{Lenarcic2013}. 
Since the excitation energy is sizable, several magnons or phonons need to be generated simultaneously in a highly non-linear process.

In the following, we assume that decay into phonons is the dominating process. The energy scale of the magnons, set by a $J$ value of a few meV \cite{Badrtdinov:2016}, is even smaller than the vibrational energy scale, and thus an even higher number of magnons as compared to phonons would be required. 
Thus, the electron-phonon coupling, expressed by the deformation potential $\lambda$, is the decisive parameter.  
Lenar{\v c}i{\v c} and co-workers \cite{Lenarcic2015} have presented a MH model coupled to a phononic Hamiltonian that allows them to estimate the lifetime of the exciton due to decay into phonons by a Zener-type expression. In their approach, 
the basic time scale for the decay is set by the ratio of the phonon frequency and the band width,  $\tau_0$. However, the actual decay rate $\tau^{-1}$ is exponentially suppressed with respect to $\tau_0^{-1}$.  
The dimensionless parameters entering this suppression factor are the order of the process, given by $n = U/(\hbar \omega_{\rm HSE})$, and the scaled electron-phonon coupling parameter $\xi = \lambda^2 / (\hbar \omega_{\rm HSE})^2$.
According to Lenar{\v c}i{\v c} and co-workers \cite{Lenarcic2015}, the expression for the lifetime has the form
\begin{eqnarray}
\tau &=& \tau_0 \sqrt{n} \exp \left( n\left( \ln \frac{n}{2\xi}-1 \right) \right) \, \mbox{with}
\label{eq:lenarch} \\
& & \tau_0 = \frac{\hbar^2 \omega_{\rm HSE}}{4 \sqrt{2 \pi} t^2 } \left(\frac{1}{2} - \frac{2 t^2}{V^2} \right)^{-1}
\nonumber
\end{eqnarray}
From our calculations, we can give a rough estimate for the phonon-related exciton lifetime using the above expression.  
For the prefactor, $\tau_0 \sim 12$~fs is obtained 
by inserting $t=25$ meV and $|t/V| \to 0$. 
Using the off-site Coulomb interaction of Ref. \onlinecite{Hansmann13}, $V=0.5$~eV would change the prefactor $\tau_0$ by less than one percent. 
Typical values for the parameters in the exponential supression factor are $n = U/(\hbar \omega_{\rm HSE}) = 68$ and $\xi \approx 10$. 
In estimating $\xi$, we used the average of the electron-phonon coupling in the 'wagging' and 'torsion' modes.
Inserting these values into eq.~(\ref{eq:lenarch}), we find that the suppression factor is very significant, of the order of $3.3 \times 10^{7}$. Thus, the basic time scale of $\tau_0=12$~fs is extended to $\tau = 400$~ns. 
In other words, if the decay into phonons is the only decay channel of the exciton in this system, 
 our calculations predict a very long lifetime which should be easily accessible in experiments. 
This finding is likely to be general for systems with strongly correlated surface states: Since the energy scales of surface phonons, typically a few tens of meV, and of the on-site Coulomb repulsion $U$, typically of the order of eV, are very different, decay of the inter-band excitation into phonons is a highly non-linear process. 
This results in a huge supression of the phononic decay channel, and a very long lifetime.  

\section{Conclusion}

Using DFT calculations with the HSE06 hybrid functional, 
the Sn/Si(111) surface with 1/3 ML coverage was found to be 
an insulating surface with bands inside the substrate band gap that can be interpreted as upper and lower Mott-Hubbard bands. The phonon spectrum and deformation potentials for electron-phonon coupling in the antiferromagnetic ground state were calculated. 
Stiffening of the wagging mode of neighboring Sn atoms is predicted from the calculations if electronic correlations are taken into account. 
Therefore, we conclude that the Jahn-Teller effect, which would be indicated by a mode softening, and the antiferromagnetic ordering of the spin system are two complementary ways of stabilizing group-IV adsorbates on Si(111). 

Vibrational spectroscopy could thus be utilized to obtain further insight into phase transitions at surfaces driven by electronic correlations. 
The Mott-Hubbard gap is found to be much larger than the typical phononic quanta; 
therefore the relaxation of electron-hole pairs into the ground state is possible only by a multi-phonon process. A very long lifetime of the excitations, in the order of hundreds of nanoseconds, is predicted 
that could be probed by two-photon photoemission spectrocopy. 

\section*{Supplemental Information}

\subsection*{Electronic band structure}

Test calculations have been performed with the VASP code \cite{VASP1,VASP2}  for the simplest case, $(\sqrt 3 \times \sqrt 3)$ unit cell with one Sn atom. 
In the non-spinpolarized PBE calculation, the surface is found to be metallic due to a Sn-induced surface state with a band width of 280 meV along $\bar \Gamma \bar{\rm M}$ and a total band width of 530 meV. 
The results are in agreement with previous studies, cf. the Supplemental Material of Li {\em et al.}, Ref.~\cite{Li13}. 

\begin{figure}[tbh]
\begin{center}
\includegraphics[width=0.45\textwidth]{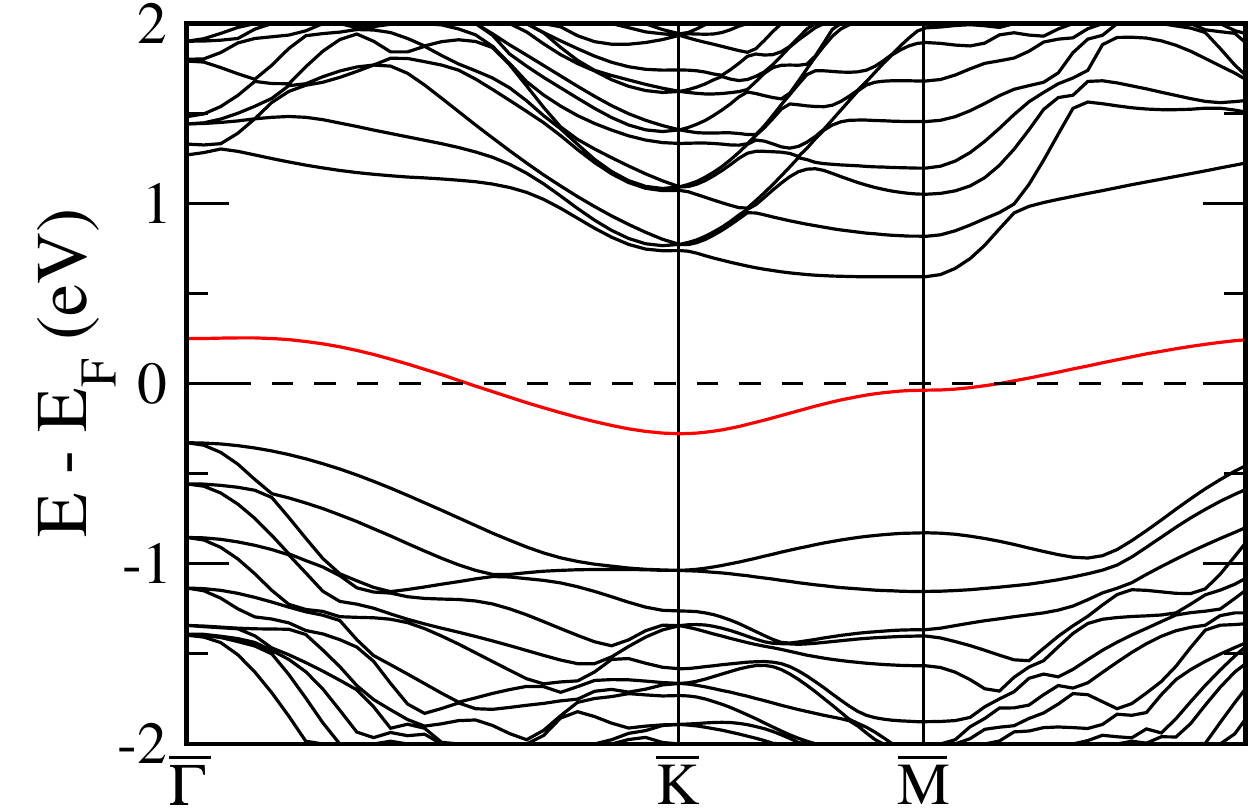}
\caption{The electronic band structure obtained in a spin-compensated calculation with the PBE functional. 
Calculations are performed in a  $(\sqrt 3 \times \sqrt 3)$ unit cell with one Sn atom.  The red band is the Sn-induced surface state with a band width of 530 meV. The position of the Fermi level is indicated by the horizontal dashed line.}
\end{center}
\end{figure}

\subsection*{Phonon band structure}
To corroborate the stiffening of the Sn wagging mode in case of AFM ordering of the magnetic moments of the Sn atoms, calculations of the phonon modes were carried out with the {\sc phonopy} package \cite{phonopy} used in conjunction with the VASP code. A supercell with $(2 \sqrt 3 \times 2 \sqrt 3)$ lateral periodicity, four Sn atoms, 120 Si atoms and 12 H atoms was constructed. Total energies and forces were calculated with the PBE functional using finite displacements of the atoms in two opposite directions for each mobile atom. The Sn atoms plus the Si atoms of the topmost three bilayers were considered to be mobile. The phonon modes resulting from diagonalizing the dynamical matrix in the {\sc phonopy} code, backfolded into the original $( \sqrt 3 \times  \sqrt 3)$ unit cell to be comparable with Figure~2 of the paper, are shown in Figure~\ref{fig:phononAFM} below. The former softening of the lowest mode near the $\bar {\rm K}$ point characteristic of the non-spinpolarized calculation has disappeared. Note that the phonon frequencies at the $\bar \Gamma$ point are slightly higher than the phonon frequencies reported in Figure~4 of the paper due to the different number of mobile atoms and different magnitude of the atomic displacements used to calculate the forces. 
 
\begin{figure}[tbh]
\begin{center}
\includegraphics[width=0.45\textwidth]{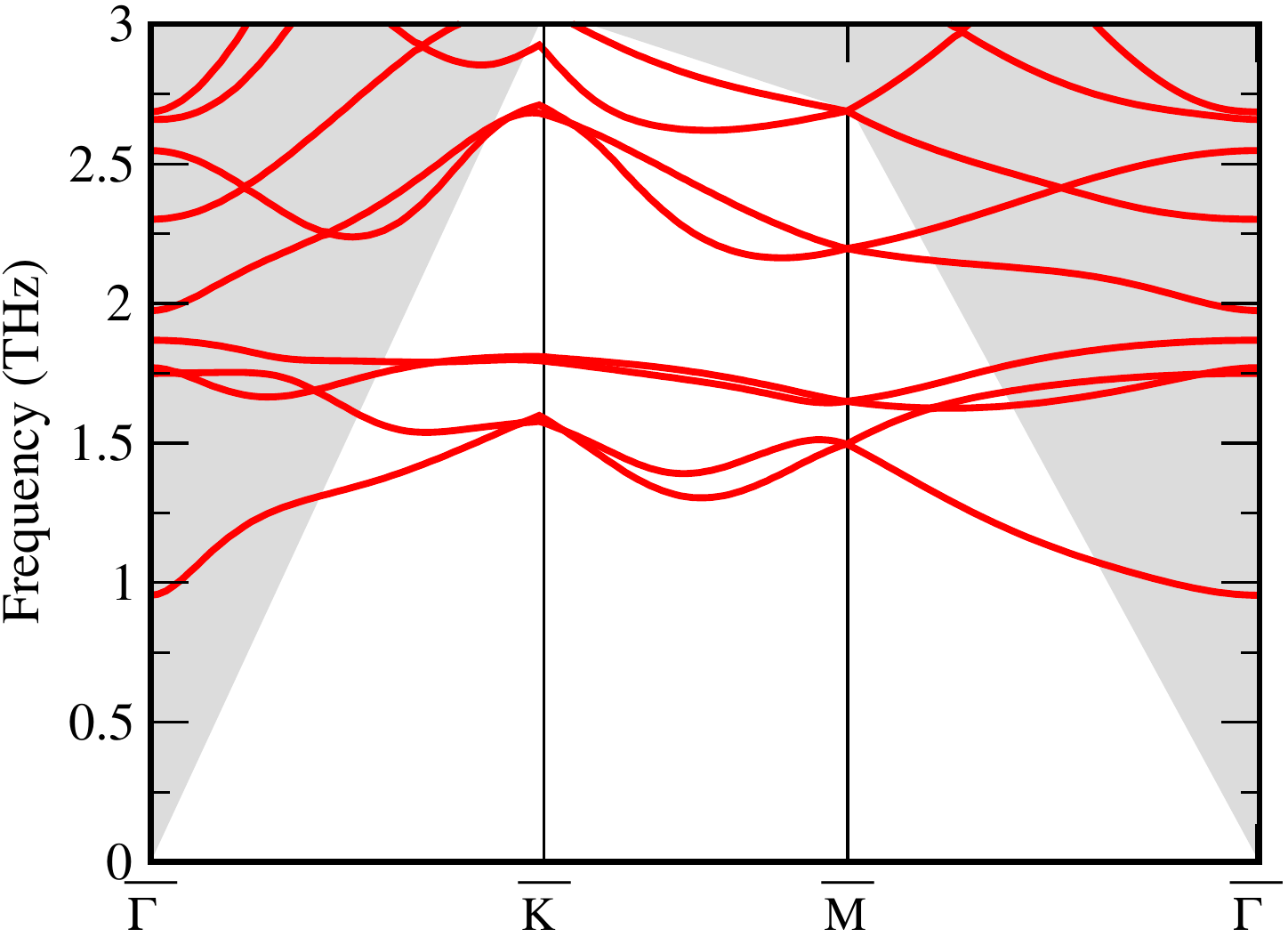}
\caption{The phonon band structure for AFM spin ordering obtained with the PBE functional along $\bar \Gamma - \bar {\rm K} - \bar {\rm M} - \bar \Gamma$ (labels refer to a  $(\sqrt 3 \times \sqrt 3)$ unit cell, cf. Fig.~1 of the paper).  The grey-shaded area indicates the region where Si-derived phonon modes are to be found.}
\label{fig:phononAFM}
\end{center}
\end{figure}

\section*{Acknowledgments}
Financial support within SFB 1242 "Non-equilibrium dynamics of condensed matter in the time domain" funded by Deutsche Forschungsgemeinschaft (DFG), project number  278162697, is gratefully acknowledged. 
We gratefully acknowledge the computing time granted by the Center for Computational Sciences and Simulation (CCSS) of the University of Duisburg-Essen and provided on the supercomputer magnitUDE (DFG Grant No. INST 20876/209-1 FUGG and INST 20876/243-1 FUGG) at the Zentrum f{\"u}r Informations-und Mediendienste (ZIM).

%


\end{document}